\documentclass[twocolumn,prl,showpacs]{revtex4}
\usepackage{graphicx}
\usepackage{dcolumn}
\usepackage{bm}

\begin{document}

\author{P.S. Corasaniti$^{(1)}$}
 
\author{B.A. Bassett$^{(2)}$}
\author{C. Ungarelli$^{(2,3)}$}
\author{E.J. Copeland$^{(1)}$}

\affiliation{(1) Centre for Theoretical Physics, University of Sussex,
   \\ Falmer, Brighton, BN1 9QJ, U.K.}

\affiliation{(2) Institute of Cosmology and Gravitation, University of Portsmouth,
\\ Mercantile House, Portsmouth, PO1 2EG, U.K. }

\affiliation{ (3) School of Physics and Astronomy, University of Birmingham, 
\\ Edgbaston, Birmingham, B15 2TT, UK.}

\title{
Model-independent dark energy differentiation with the ISW effect
}

\begin{abstract}
We study the integrated Sachs-Wolfe effect using a model-independent
parameterization of the dark energy equation of state, w(z). Cosmic
variance  severely restricts the
class of models distinguishable from $\Lambda$CDM unless
w(z) currently satisfies $w_Q^o > -0.8$, or
exhibits  a rapid, late-time, transition  at redshifts $z<3$. Due to
the degeneracy with other cosmological parameters, models with a
slowly varying w(z) cannot be differentiated from each other or from a
cosmological constant. This may place a fundamental limit
on our understanding of the origin of the presently observed acceleration.
\end{abstract}

\maketitle

\underline{\em 1)  Introduction} --
Cosmological distance measurements of high redshift type Ia supernova, combined
with measurements of the cosmological volume via galaxy cluster surveys, weak
lensing tomography and quasar clustering
can provide a new insight into the nature of the dark energy sourcing the present
acceleration of the universe. In spite of their potential for breaking 
degeneracies between
cosmological parameters such tests are limited by our ignorance of effects
such as possible supernova evolution or nonlinear galaxy cluster physics.
In this sense the cosmic microwave background (CMB) observations are uniquely
pure, limited only by systematic experimental effects  
and cosmic variance (an unavoidable theoretical error). This purity 
motivates us to ask the following question:
how much model-independent information can the CMB
give us about the dark energy ?
We will show that if the  
dark energy only clusters on very large scales then only models 
with equations of state, $w(z)$, which vary 
rapidly at low redshifts $z < 3$ can be distinguished  from the `concordance'
$\Lambda$CDM model using the CMB alone. 
The importance of differentiating between a 
cosmological constant ($\Lambda$) and  
dynamical models of dark energy such as one (or more) evolving 
scalar fields (`Quintessence' - Q) \citep{WETT,RATRA,STEIN} 
can hardly be overstated. $\Lambda$ is essentially
anti-gravity, the other is a new long-range force. 
We tackle this differentiation issue using a very general parameterization of 
$w(z) \equiv p/\rho$  described in \citep{CORAS2} and used in 
simplified form in \citep{BASS}. This form accurately encompasses 
most quintessence models, independent of the number of fields involved, 
and many other models for the acceleration.
In principle cosmological distance measurements
can distinguish between $\Lambda$CDM and QCDM models by constraining the
present value of the dark energy equation of state $w_Q^o$.
This does not work however, since current CMB data suggest that  
$w_Q^o$ is close to the 
cosmological constant value of $-1$ \citep{CORASA,BACCI,BEAN,BASS,HANN}.
However, the situation is not hopeless since dark energy clusters on
very large scales and hence can leave a distinctive contribution in the
anisotropies of the Cosmic Microwave
Background radiation (CMB).  This is particularly clear in the case 
of a rapid late-time transition in $w(z)$. Indeed current data does (weakly)
prefer exactly
such a situation \citep{BASS}. \\
In this {\em letter} we focus on the relation between the dark 
energy properties
and the integrated Sachs-Wolfe effect (ISW) in the CMB power spectrum.
Although this is model dependent
we have recently shown that an appropriate parameterization of
the equation of state, $w(z)$, accurately describes a  
large class of models given by eq. (4) in ref. \citep{CORAS2}.
Assuming that the dark energy contribution during the radiation era is 
negligible, then different dark energy models
are specified by the vector $\overline{W} = (w_Q^o, w_Q^m, a_c^m, \Delta)$
which respectively specifies the equation of state today (o) and 
during the matter era (m); 
the scale factor where the equation of state changes from $w_Q^m$ to $w_Q^o$
and the width of the transition.
We now discuss how these parameters affect the CMB and show that 
only a small range of $\overline{W}$s
leave an imprint on the CMB distinguishable from $\Lambda$CDM.
\\
\underline{\em 2) ISW effect in dark energy models} --  
A late-time mechanism to generate anisotropies is due to CMB photons
climbing in and out of evolving gravitational potentials \citep{REES}.
During the matter dominated era the gravitational potential
$\Phi$ associated
with the density perturbations
is constant and there is no ISW effect. However in $\Lambda$CDM models
$\Phi$ starts decaying at redshifts when $\Lambda$
starts
to dominate, producing large angular scale anisotropies \citep{CRI}.
In dark energy scenarios the cosmic acceleration is not the
only contribution to the decay of $\Phi$:
on large scales the clustering properties alter
the growth rate of matter perturbations \citep{CALD,MA2}.
It is the signal of this clustering \cite{HU0} that we are hunting 
in as model-indepenent a way as possible. 
We assume a flat spatial geometry and fix the value of the Hubble constant
$H_o=70$ Km$s^{-1}$$Mpc^{-1}$, the scalar spectral index $n=1$,
the baryon density $\Omega_b h^2=0.021$
and the amount of matter (CDM) $\Omega_{m}=0.3$.
We can usefully distinguish two classes of models, 
\begin{itemize}
\item first those with a 
slowly varying equation of state for which $0 < a_c^{m}/ \Delta < 1$, as in
the case of the inverse power law potential \citep{ZLATEV}, and second: 
\item a rapidly
varying $w(z)$, such as the 'Albrecht-Skordis` model \citep{ALB} and
the two exponential potential \citep{BARRE}, with 
$ a_c^m/ \Delta > 1$. This class also includes many interesting 
radical models such as vacuum metamorphosis \citep{PARK}, 
late-time phase transitions \citep{HILL}, and backreaction-induced 
acceleration \cite{WETT2}.
\end{itemize}
We show these two classes in fig.1.
The red solid line corresponds to dark energy that tracks
the dust during the matter era ($w_Q^m=0.0$) and 
evolves slowly toward $w_Q^o=-1$, and the 
blue dotted line corresponds to a model with a rapid transition in its equation
of state at $a_c^m=0.1$ ($z = 9$). 
Given current data it is worth studying the case with $w_Q^o=-1$, 
(since it is also the most difficult to di
stinguish from $\Lambda$CDM) 
whilst allowing the other parameters $w_Q^m$ and $a_c^m$, to vary.

\begin{figure}[ht]
\includegraphics[height=6cm,width=6cm]{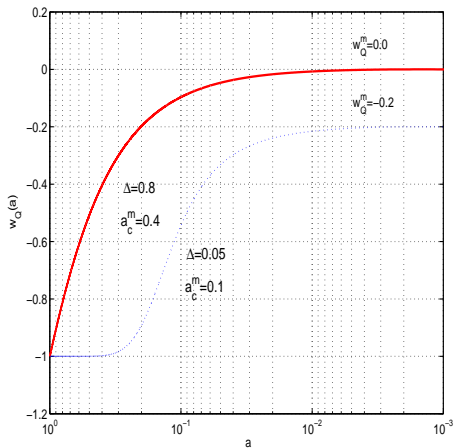}
\caption{ Time evolution of the equation of state
for two classes of models, with slow (red solid line) and rapid transition 
(blue dotted line). The dark energy parameters specify the features of $w_Q(a)$}.
\end{figure}
Fig.2 shows the anisotropy power spectrum, $C_l^{ISW}$, produced through
the integrated Sachs-Wolfe effect by  a rapidly evolving ($top$ $panels$) and 
a slowly evolving ($bottom$ $panels$) equation of
state; the red (solid) line corresponds to the $\Lambda$CDM model.
As we can see in the top left panel (fig.2a), varying $a_c^m$ can 
produce a strong
ISW. The effect is larger if the transition in the equation of state occurs
at redshifts $z<3$. On the other hand the $C_l^{ISW}$ is the same as in 
cosmological constant regime if $a_c^m<0.2$ ($z>4$). 
In the top right panel (fig.2b)
we plot the ISW for two different values of $w_Q^m$, corresponding to
$w_Q^m=0.0$ (dashed line) and $w_Q^m=-0.1$ (dot-dash line).
We note that the signal is larger if the quintessence field is perfectly
tracking the background component. But as $w_Q^m$ diverges from the dust
value the ISW effect becomes the same as in $\Lambda$CDM. 
This means that even
for rapidly varying $w(z)$ (small $\Delta$), the ISW is distinguishable 
from that in the $\Lambda$CDM scenario only if $w(z)$ 
during matter domination closely mimics the dust value and the 
transition occurs at low redshifts, $z<3$. 
We can see the amplitude of the integrated Sachs-Wolfe effect is smaller in
slowly varying models ($bottom$ $panels$). As we expect the $C_l^{ISW}$
is independent of $a_c^m$ (fig.2c), since for these models a
different value in the transition redshift does not produce a large 
effect on the evolution of the dark energy density.
In fig.2d, the ISW power spectrum is large for $w_Q^m=0.0$ (dash line)
and becomes smaller than the cosmological constant on horizons
scales as $w_Q^m$ has negative values (dot-dashed line),
and increases toward $\Lambda$ for $w_Q^m$ approaching $-1$. 
This class of models is then more difficult to distinguish 
from the $\Lambda$CDM if the equation of state today is close to 
$w_{\Lambda} = -1$.
This can be qualitatively explained noting that perfect 
tracking between dark energy and CDM causes a delay in the time when the 
gravitational potential starts to decay, compared to the case of $\Lambda$CDM. 
This effect is stronger for models with rapidly varying equation 
of state since the rapid change in $w_{Q}$ produces a stronger variation in 
the gravitational potential.
\begin{figure}[ht]
\includegraphics[height=8cm,width=8cm]{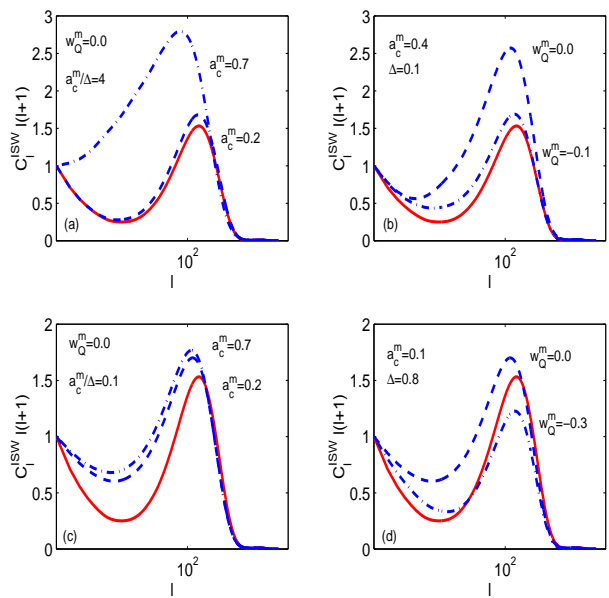}
\caption{Power spectrum of the ISW for rapidly varying models (top panels) and
slowly varying ones (bottom panels). The solid red line is the ISW effects produced
in the cosmological constant case. Detailed explanation in the text.} 
\end{figure}
\begin{figure}[ht]
\includegraphics[height=8cm,width=8cm]{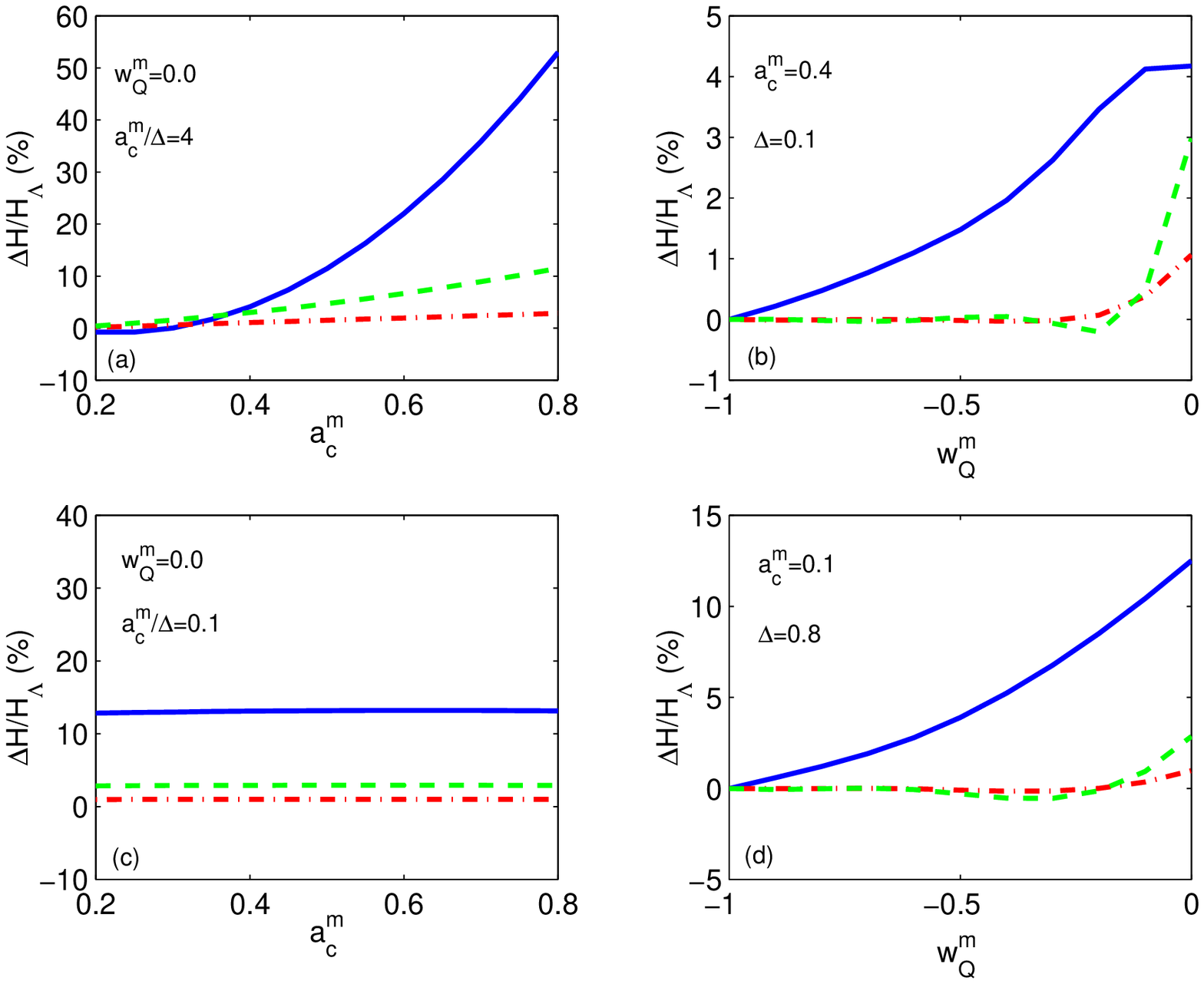}
\caption{Relative difference of $H_1$ (blue solid line), $H_2$ (green
dash
line) and $H_3$ (red dash-dot line) to the $\Lambda$CDM model, for rapidly
varying models ($top$ $panels$), and with slow transition ($bottom$ $panels$).
For these models the present value of the equation of state is $w_Q^o=-1$.} 
\end{figure}
\begin{figure}[ht]
\includegraphics[height=8cm,width=8cm]{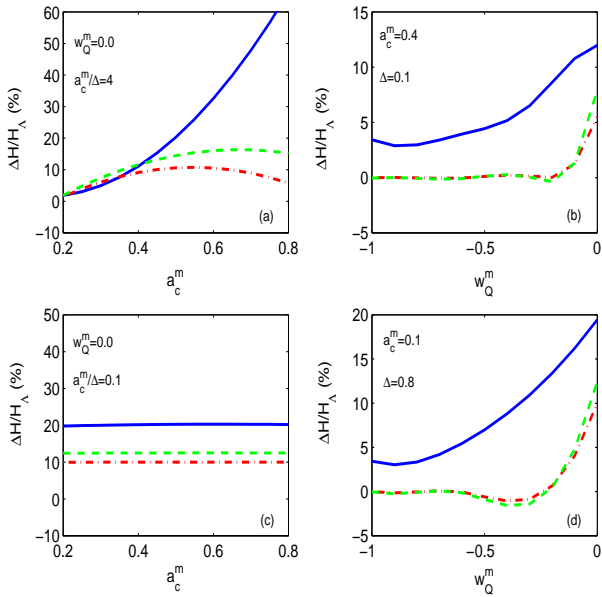}
\caption{As in fig.3 for $w_Q^o=-0.88$.} 
\end{figure}

\underline{\em 3) CMB power spectrum} -- The imprint of the ISW effect
in the CMB spectrum results in a boost in
power at low multipoles. This affects the position of the first acoustic 
peak and  the COBE normalization. In particular, since some of the
anisotropies at large scales are produced at late times,
as we have seen in the previous section,
normalizing the spectrum relative to the COBE measurements produces a suppression of power
at smaller scales. 
A simple way to characterize the amplitude of 
the CMB spectrum is to consider the height of the first three acoustic peaks 
relative to the power at $l=10$, i.e. $H_i=C_{l_i}/C_{10}$, and their
multipole positions $\ell_i$, $i = 1,2,3$
\citep{HU}.
These numbers allows us to quantify the
discrepancy between the dark energy models and $\Lambda$CDM. 
In particular since $H_1$ depends on the amplitude of the power spectrum
at the multipoles characteristic of the ISW effect, we expect $H_1$ to
be more sensitive to Quintessence signatures.
We have computed the CMB spectra for the class of models with $w_Q^o=-1$
described in the previous section, determined the parameters, 
$H_i$, (plotted in fig.3) and compared them with those of the $\Lambda$CDM 
spectrum.
The rapidly varying models are shown in the top panels. 
We can see the strong ISW effects produced by changing $a_c^m$ are now
evident in the large discrepancy between $H_1$ and $H_1^{\Lambda}$
(blue line) (fig.3a): it can be larger then 20 per cent for $a_c^m>0.6$.
The effect on $H_2$, $H_3$ is smaller. However varying $w_Q^m$ (fig.3b)
produces a discrepancy of only order $4$ per cent on $H_1$, while
$H_2$ and $H_3$ remain the same as in $\Lambda$CDM.
For a slowly varying equation of state, $H_1$, $H_2$ and $H_3$ are 
independent of $a_c^m$ (fig.3c). The dark energy imprint
is only on $H_1$ for which the discrepancy to the $\Lambda$ case
is about 10 per cent. Such discrepancy decreases
when changing the value of the quintessence equation of
state during matter from $w_Q^m=0.0$ to $w_Q^m=-1$ (fig.3d). 
Values of the equation of state
today $w_Q^o > -1$ imply a stronger ISW effect.
Consequently the curves of fig.3 are shifted upwards. For instance in fig.4
we plot the class of models previously analysed, with $w_Q^o=-0.88$. We note
the same behavior as we vary the dark energy parameters, but the discrepancy
with the $\Lambda$CDM model is now larger. 
In fig.4d it is worth noticing
the case $w_Q^m=-1$, that corresponds to a model very similar to 
a `k-essence' model \cite{MUKA}.
We can see that the relative difference with the $\Lambda$CDM case 
is of the order of a few percent, in agreement with \cite{ERIC} 
for the same value of $w_Q^o=-0.88$.
At this point we ask the key question whether such differences are observable. 
We have shown that $H_1$ is a good estimator of the ISW effect, and that it
is a tracer of the dark energy imprint on the
CMB. However its estimation from the data
will be affected by cosmic variance at $l=10$. Hence with even perfect 
measurements of the first acoustic peak the
uncertainty on $H_1$ will be dominated by the  
$30$ per cent uncertainty due to cosmic variance. 
With the plots of fig.3 and fig.4 in mind, this means that if the present value
of the equation of state is close to $-1$,
slowly varying dark energy models are hardly distinguishable from $\Lambda$CDM,
while rapidly 
varying ones can produce a detectable signature only if the transition in the
equation occurred at $a_c^m>0.7$, but in any case it will be difficult to 
constrain $w_Q^m$. The degeneracy of $H_1$ with 
the baryon density is marginal since varying $\Omega_b h^2$ mainly 
affects $H_2$ which is insensitive to dark energy effects.
Hence only an accurate determination of the angular
diameter distance, inferred from the location of the acoustic peaks,
would allow detection of such deviations from the cosmological
constant model. The relation between the position of the CMB peaks and the dark
energy has been widely discussed (\citep{DORAN2} and references 
therein). The shift of the multipole 
positions ($\ell_i$) of the acoustic peaks caused by the 
evolution of the dark energy in the class of models analysed in fig.3 can be
seen in fig.5, where we plot the 
relative difference of $l_1, l_2$ and $l_3$ to the $\Lambda$ case.
We note that due to the additional shift induced on the first acoustic peak
by the ISW effect the difference with the $\Lambda$CDM model 
for the first peak is generally larger than for the second and third peaks. 
As with the comparison of the amplitude of the CMB spectrum,
the largest effect is produced by models with a rapid transition occurring at
small redshifts.
However the degeneracy of the angular diameter distance,
in particular with the value of
Hubble constant and the amount of dark energy density, will limit our ability
to put tight constraints on the
dark energy parameters. Fortunately there are 
alternative ways in which  these problems can be alleviated,
for instance, cross-correlating
the ISW  effect with the large scale structure of
the local universe \citep{HUW,CO,BOUGHN}.
An efficient approach would also be to
combine different observations in order to break the
degeneracies with the cosmological parameters \citep{WAS,FRIEMAN}.

\begin{figure}[ht]
\includegraphics[height=8cm,width=8cm]{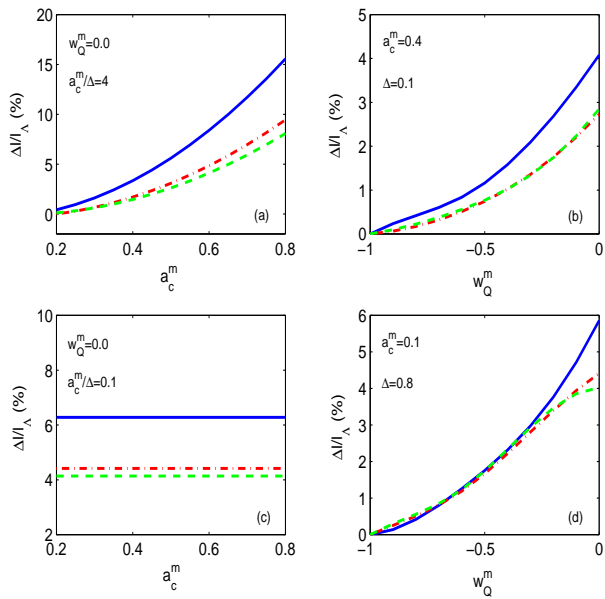}
\caption{Relative difference of $l_1$ (blue solid line), $l_2$ (green
dashed
line) and $l_3$ (red dot-dashed line) to the $\Lambda$CDM model, for rapidly
varying models ($top$ $panels$), and with slow transition ($bottom$ $panels$).
For these models the present value of the equation of state is $w_Q^o=-1$.} 
\end{figure}

\underline{\em 4) Conclusions} -- The next generation of 
high resolution CMB experiments will measure
the anisotropy power spectrum with an accuracy close to theoretical limits.
It is therefore of particular interest to study the sensitivity
of CMB observations to the effects produced by a
dark energy component in the CMB. 
A quintessential contribution leaves a distinctive signature in the
ISW effect. On the other hand such an imprint occurs at low
multipoles, consequently cosmic variance strongly limits the possibility 
of differentiating dark energy models from a cosmological
constant.
In particular we find that for values of $-1 \leq w_Q^o<-0.8$,
the only distinguishable cases are those with a rapidly varying 
equation of state. In fact for the slowly varying models, the ISW
is the same as in $\Lambda$CDM and
a deviation from $w_Q^o=-1$ can be inferred only
from an accurate determination of the location of the acoustic peaks.
However such measurements will be affected by the degeneracy of the angular
diameter distance with the value of the Hubble constant and the amount
of dark energy density. 
Therefore CMB observations are insensitive
to this class of models and we will be left with a fundamental uncertainty 
as to whether the cosmic acceleration is due to an evolving field or 
a cosmological constant, an issue of great theoretical importance 
for our understanding of the foundations of quantum gravity and string theory. 

\begin{acknowledgements}
We thank Rob Lopez for discussions and supplying part of 
the code upon which this  work is based. \\

PSC is supported by a  University of Sussex bursary. The research of BB and CU 
was supported by PPARC  grant PPA/G/S/2000/00115.
\end{acknowledgements}

\end{document}